\documentclass[12pt]{article}
\usepackage{geometry}
\usepackage{amsmath,amssymb}
\usepackage[nosort]{cite}
\usepackage{float}
\usepackage{latexsym}
\usepackage{graphicx}
\usepackage{color}
\usepackage{subfig}
\usepackage{color}

\newcommand{\beq}{\begin{equation}}
\newcommand{\be}{\begin{equation}}
\newcommand{\ee}{\end{equation}}
\newcommand{\bea}{\begin{eqnarray}}
\newcommand{\eea}{\end{eqnarray}}

\newcommand{\pa}{\partial}

\newcommand{\nn}{\nonumber}

\newcommand{\ma}{{\mathcal{A}}}
\newcommand{\mf}{{\mathcal{F}}}

\newcommand{\mc}{{\mathcal{C}}}

\begin{document}

\begin{titlepage}
\hbox to \hsize{\hspace*{0 cm}\hbox{\tt }\hss
    \hbox{\small{\tt }}}

\vspace{1 cm}

\centerline{\bf \Large Thermodynamics of black branes}

\vspace{.6cm}

\centerline{\bf \Large in asymptotically Lifshitz spacetimes }

%\vspace{.6cm} \centerline{\bf \Large }
\vspace{1 cm}
 \centerline{\large  $^\dagger\!\!$ Gaetano Bertoldi, $^\dagger\!\!$ Benjamin A. Burrington and $^\dagger\!\!$ Amanda W. Peet}

\vspace{0.5cm}
\centerline{\it ${}^\dagger$Department of Physics,}
\centerline{\it University of Toronto,}
\centerline{\it Toronto, Ontario, Canada M5S 1A7. }

\vspace{0.3 cm}

\begin{abstract}
Recently, a class of gravitational backgrounds in $3+1$ dimensions have been proposed as holographic duals to a Lifshitz theory describing critical phenomena in $2+1$ dimensions with critical exponent $z\geq 1$.  We continue our earlier work \cite{Bertoldi:2009vn}, exploring the thermodynamic properties of the ``black brane'' solutions with horizon topology $\mathbb{R}^2$.  We find that the black branes satisfy the relation $\mathcal{E}=\frac{2}{2+z}Ts$ where $\mathcal{E}$ is the energy density, $T$ is the temperature, and $s$ is the entropy density.  This matches the expected behavior for a $2+1$ dimensional theory with a scaling symmetry  $(x_1,x_2)\rightarrow \lambda (x_1,x_2)$, $t\rightarrow \lambda^z t$.
\end{abstract}
\end{titlepage}

\section{Introduction}
Since the Maldacena conjecture \cite{Maldacena:1997re}, holography has offered an interesting new tool to explore strongly coupled field theories (for a review, see \cite{Aharony:1999ti}).  In this framework, black hole backgrounds are dual to strongly coupled plasmas, and using these backgrounds, one can extract hydrodynamic and thermodynamic properties of the plasma \cite{Gubser:1996de2,Gubser:1996de,Witten:1998zw}.

Recently, much effort has gone into describing quantum critical behavior in condensed matter systems using holographic techniques (for a review, see \cite{Hartnoll:2009sz}).  Quantum critical systems exhibit a scaling symmetry
\be
t\rightarrow \lambda^{z}t, \quad x_i\rightarrow \lambda x_i
\ee
similar to the scaling invariance of pure AdS ($z=1$) in the Poincar\'e patch.  From a holographic standpoint, this suggests the form of the spacetime metric
\be
ds^2=L^2\left(r^{2z} dt^2+ r^2 dx^i dx^j \delta_{i j}+\frac{dr^2}{r^2}\right), \label{scalemetric}
\ee
where the above scaling is realized as an isometry of the metric along with $r\rightarrow \lambda^{-1} r$ (for our purposes, $i=1,2$).
Other metrics exist with the above scaling symmetry, but also with an added Galilean boost symmetry \cite{Son:2008ye,Balasubramanian:2008dm,Mazzucato:2008tr} which we will not consider here (black brane solutions in these backgrounds were discussed in \cite{Yamada:2008if}).  There has also been some success at embedding a related metric into string theory \cite{Azeyanagi:2009pr} with anisotropic (in space) scale invariance.  This may serve as a template for embedding metrics of the form (\ref{scalemetric}) into string theory.  Here, however, we will continue our study of the model in \cite{Kachru:2008yh} (some analysis of generalizations of this model appear in \cite{Taylor:2008tg,Pang:2009ad,Kovtun:2008qy}).
The authors of \cite{Kachru:2008yh} constructed a 4D action that admits the metric (\ref{scalemetric}) as a solution\footnote{This background was earlier studied in \cite{Koroteev:2007yp}, however, without an action principle that admits the above metric as a solution.}, which is equivalent to the action \cite{Taylor:2008tg}
\be
S=\frac{1}{16 \pi G_4}\int d^4x\sqrt{-g}\left(R-2\Lambda-\frac14 \mf_{\mu \nu} \mf^{\mu \nu}-\frac{c^2}{2}\ma_{\mu}\ma^{\mu}\right)\label{staction}
\ee
($\mathcal{F}=d\mathcal{A}$) with terms in the action parameterized by
\be
c=\frac{\sqrt{2{Z}}}{\hat{L}}, \quad \Lambda=-\frac12\frac{Z^2+Z+4}{\hat{L}^2} \label{lifsolcons}.
\ee
The solution discussed in \cite{Kachru:2008yh} has the metric (\ref{scalemetric}) and
\bea
\ma&=& L^2\frac{r^z}{z}\sqrt{\frac{2z(z-1)}{L^2}}dt \label{lifsol}
\eea
with the identification $z=Z$ and $L=\hat{L}$ or the identification $z=4/Z, L=\frac{2}{Z}\hat{L}$ and is defined for $z\geq 1$ (with $z=1$ giving AdS$_4$).

In our current work, we will analyze the thermodynamics of black brane solutions which asymptote to (\ref{scalemetric}) \footnote{We constructed numeric solutions for this system in \cite{Bertoldi:2009vn} (seen as the large $r_0$ limit of this earlier work), however, our current work does not depend on any numeric analysis.}.  These black brane solutions have been studied numerically in \cite{Bertoldi:2009vn,Danielsson:2008gi,Mann:2009yx}.  However, we will show that their energy density $\mathcal{E}$, entropy density $s$ and temperature $T$ are related by
\be
\mathcal{E}=\frac{2}{2+z}Ts \label{thermorel2}
\ee
using purely analytic methods.

In the following section we will introduce the ingredients needed to prove (\ref{thermorel2}).  Particularly important is the existence of a conserved quantity, used to relate horizon data to boundary data.  In the final section, we combine these ingredients into the result (\ref{thermorel2}).  We also show that this relation is expected in the dual field theory as a result of the scaling symmetry $(x_1,x_2)\rightarrow \lambda (x_1,x_2)$, $t\rightarrow \lambda^z t$.

\section{Summary of earlier results}
In our earlier work \cite{Bertoldi:2009vn}, the action (\ref{staction}) was reduced on the Ansatz
\bea
ds^2&=&-e^{2A(r)}dt^2+e^{2B(r)}((dx^1)^2+ (dx^2)^2)+e^{2C(r)}dr^2 \nn \\
&&\ma=e^{G(r)}dt
\eea
to give a one dimensional Lagrangian
\bea
L_{1D}&=&4e^{(2B+A-C)}\pa B \pa A+2e^{(2B+A-C)} (\pa B)^2+\frac12 e^{(-A+2B-C+2G)}(\pa G)^2 \nn \\
&&-2\Lambda e^{(A+2B+C)}+\frac12 c^2 e^{(-A+2B+C+2G)}. \label{actions}
\eea
The equations of motion following from this action have solutions given by (\ref{scalemetric}), (\ref{lifsol}).  Further, there are the known black brane solutions that asymptote to AdS$_4$,
\bea
ds^2&=& \left(\frac{-3}{\Lambda}\right)\left(-r^2 f(r)dt^2+r^2(dx_1^2+dx_2^2)+\frac{dr^2}{r^2f(r)}\right), \nn \\
f(r)&=&1-\frac{r_0^3}{r^3} \nn \\
\ma&=&0.
\eea
For the remainder of the paper, we will be concerned with black branes which asymptote to the Lifshitz background (\ref{scalemetric}), (\ref{lifsol}) for $z \neq 1$.

In \cite{Bertoldi:2009vn}, a Noether charge was found which is associated with the shift
\be
\begin{pmatrix}
A(r) \\
B(r) \\
C(r) \\
G(r) \\
\end{pmatrix}
\rightarrow
\begin{pmatrix}
A(r)+ \delta \\
B(r)- \frac{\delta}{2}\\
C(r)+ 0 \\
G(r)+ \delta  \\
\end{pmatrix}
\ee
with $\delta$ a constant.  The above represents a diffeomorphism which preserves the volume element $dt dx^1 dx^2$.  This is why it is inherited as a Noether symmetry in the reduced Lagrangian.  The associated conserved quantity is
\be
(2e^{(A+2B-C)}\pa A- 2e^{(A+2B-C)}\pa B-e^{(-A+2B-C+2G)} \pa G)\equiv D_0 \label{D0}.
\ee

\subsection{The perturbed solution near the horizon}

We begin by first reviewing the results found in \cite{Bertoldi:2009vn} for the expansion near the horizon.  We require that $e^{2A}$ goes to zero linearly, $e^{2C}$ has a simple pole, and $e^{G}$ goes to zero linearly to make the flux $d\ma$ go to a constant (in a local frame or not).  Further, we take the gauge $B(r)=\ln(Lr)$ for this section.  We expand
\bea
&&A(r)=\ln\left(r^z L\left(a_0(r-r_0)^{\frac12}+a_0a_1(r-r_0)^{\frac32}+\cdots\right)\right), \qquad B(r)=\ln(rL) \nn \\
&&C(r)= \ln\left(\frac{L}{r}\left(c_0(r-r_0)^{-\frac12}+c_1(r-r_0)^{\frac12}+\cdots\right)\right), \\ &&G(r)=\ln\left(\frac{L^2r^z}{z}\sqrt{\frac{2z(z-1)}{L^2}}\left(a_0g_0(r-r_0)+a_0g_1(r-r_0)^2+\cdots\right)\right). \nn
\eea
Note that by scaling time we can adjust the constant $a_0$ by an overall multiplicative factor (note the use of $a_0$ in the expansion of $G(r)$ as well, as $e^{G}$ multiplies $dt$ for the one-form $\ma$).  We will need to use this to fix the asymptotic value of $A(r)$ to be exactly $\ln(r^zL)$ with no multiplicative factor inside the log.

We plug this expansion into the equations of motion arising from (\ref{actions}), and solve for the various coefficients.  We find a constraint on the 0th order constants: as expected not all boundary conditions are allowed.  We solve for $c_0$ in terms of the other $g_0$ and $r_0$, and find
\bea
c_0 = \frac{\sqrt{(2z+g_0^2 r_0(z-1))} r_0^{\frac32}}{\sqrt{z}\sqrt{{(z^2+z+4)r_0^2}}}. \label{c0equ}
\eea
All further coefficients (e.g. $g_1$) are determined from the two constants $r_0$ and $g_0$.

We evaluate (\ref{D0}) at $r=r_0$ using the above expansion to find
\be
D_0=\frac{r_0^{z+3} L^2 a_0}{c_0}. \label{D0r0}
\ee
This must be preserved along the flow in $r$.  We will use this to relate constants at the horizon to coefficients that appear in the expansion at $r=\infty$.

\subsection{The perturbed solution near $r=\infty$}
\label{rinftysect}
We now turn to the question of the deformation space around the solution given in (\ref{lifsol}) and (\ref{lifsolcons}).  We take the expansion of the functions
\bea
&&A(r)=\ln(r^z L)+\epsilon A_1(r), \qquad B(r)=\ln(rL)+\epsilon B_1(r) \nn \\
&&C(r)= \ln\left(\frac{L}{r}\right)+\epsilon C_1(r), \qquad G(r)=\ln\left(\frac{L^2r^z}{z}\sqrt{\frac{2z(z-1)}{L^2}}\right) +\epsilon G_1(r).
\eea
Using straightforward perturbation theory, we may find the solutions in the $B_1=0$ gauge
\bea
A_1(r)&=& {\mathcal{C}}_0 \frac{(z-1)(z-2)}{(z+2)} r^{-z-2} +\mc_2\left(z^2+3z+2-(z+1)\gamma\right)r^{-\frac{z}{2}-1-\frac{\gamma}{2}}
\label{A1expansion} \\
B_1(r)&=&0  \label{B1expansion}\\
C_1(r)&=& -{\mathcal{C}}_0(z-1)r^{-z-2} +\mc_2\left(z^2-7z+6-(z-1)\gamma\right)r^{-\frac{z}{2}-1-\frac{\gamma}{2}} \label{C1expansion} \\
G_1(r)&=& {\mathcal{C}}_0 \frac{2(z^2+2)}{z+2}r^{-z-2}+\mc_24z(z+1)r^{-\frac{z}{2}-1-\frac{\gamma}{2}} \label{G1expansion}
\eea
where we have defined the useful constant
\be
\gamma=\sqrt{9z^2-20z+20}.
\ee
In the above, we have dropped certain terms in the expansion from \cite{Bertoldi:2009vn}.  We have dropped them so that we meet the criterion of \cite{Ross:2009ar} to be sufficiently close to the Lifshitz background (\ref{scalemetric}), (\ref{lifsol}).  We may evaluate the conserved quantity, and we find
\be
D_0=-\frac{2(z-1)(z-2)(z+2)L^2}{z} \mc_0. \label{D0infty}
\ee
One may worry that nonlinearities may contribute to the value of this constant.  However, one may examine the powers of $r$ available in \cite{Bertoldi:2009vn}, and quickly be convinced that the higher nonlinear contributions will be zero once we meet the criterion of \cite{Ross:2009ar} (i.e. dropping the ``bad'' modes).  Hence, the above constant is the value of $D_0$ throughout the flow in $r$.

\subsection{Gauge invariance}
\label{gaugesection}
In the previous sections, we have gauge fixed by taking $B_1(r)=0$.  Here, we write down the linearized gauge transformations  that will allow us to switch to other gauges in perturbation theory (used near $r=\infty$).  The transformation
\bea
A_1(r)&&\rightarrow A_1(r)+\frac{z}{r}\delta(r), \quad B_1(r)\rightarrow B_1(r)+\frac{1}{r}\delta(r)  \label{gaugesym}\\
C_1(r)&&\rightarrow C_1(r)+r\pa_r\left(\frac{\delta(r)}{r}\right), \quad G_1(r)\rightarrow G_1(r)+\frac{z}{r}\delta(r) \nn
\eea
corresponds to infinitesimal coordinate transformations $r\rightarrow r+ \epsilon \delta(r)$.  One can see that such a shift leaves the equations invariant to leading order in $\epsilon$ when expanding about the solution (\ref{scalemetric}),  (\ref{lifsol}).

\section{The black brane thermodynamics}

In the above, we have calculated the conserved quantity $D_0$ in two regions: near the horizon $r=r_0$ and in the asymptotic region $r=\infty$.  We may use this to solve for $\mc_0$ in terms of the horizon data
\be
\mc_0=-\frac12 \frac{r_0^{z+3}a_0 z}{c_0(z-1)(z-2)(z+2)}.
\ee
We will see that $\mc_0$ is proportional to the energy density $\mathcal{E}$ of the background.

Indeed, the authors of \cite{Ross:2009ar} identified the energy density of the background in terms of the coefficient of the $r^{-z-2}$ term in the expansion at infinity for any $z$.  This mode was also identified in \cite{Bertoldi:2009vn} as the mass mode using background subtraction. However, for $1\leq z \leq 2$ there were additional divergences that were not canceled.  These were cured using the local counter terms of \cite{Ross:2009ar}.

One may not, however, directly use the results of \cite{Ross:2009ar} because of a different choice of gauge: above we use $B_1(r)=0$ and the authors of \cite{Ross:2009ar} use $C_1(r)=0$.  We may easily switch to this gauge by taking
\be
\delta(r)=-r\int{\frac{C_1(r)}{r}} dr
\ee
and transforming the other fields appropriately (near $r=\infty$).  In the above integral, we make sure to take the constant of integration so that at large $r$ the correction fields vanish.  We find that in the $C_1(r)=0$ gauge
\be
A_1(r)=- 2\frac{(z-1)\mc_0 r^{(-z-2)}}{(2+z)}+\cdots
\ee
where $\cdots$ are the other terms we are not concerned about.  This allows us to compare our results directly to the calculation of \cite{Ross:2009ar}.  We identify $2A_1(r)=f(r)$ where $f(r)$ appears in equation (5.27) \cite{Ross:2009ar}.  Therefore, we identify
\be
c_{1,{\rm RS}}=-(z-1)\mc_0
\ee
where $c_{1,{\rm RS}}$ appears in \cite{Ross:2009ar} as the coefficient of $r^{-z-2}$.  Therefore, equation (5.31) of \cite{Ross:2009ar} becomes
\be
\mathcal{E}= -\frac{4(z-2)(z-1)}{z}\mc_0=\frac{2 r_0^{z+3}a_0}{(2+z)c_0}.
\ee
where $\mathcal{E}$ is the energy density.  However, we should note that this is a unitless energy.  Restoring units we find
\be
\mathcal{E}=\frac{2 r_0^{z+3}a_0}{(2+z)c_0}\frac{1}{16 \pi G_4 L}.
\ee
From the metric, it is easy to read off the area of the horizon, and therefore the entropy density
\be
s=\frac{4\pi r_0^2}{16 \pi G_4}.
\ee
Further, one can easily read off the temperature from the expansion at the horizon \cite{Bertoldi:2009vn}
\be
T=\frac{r_0^{z+1}a_0}{4 \pi c_0 L}.
\ee
From this, we may read an interesting thermodynamic relationship
\be
\mathcal{E}=\frac{2}{2+z}Ts, \label{thermoRel}
\ee
which is the main result of our work.

We compare this expression with the known $z=1$ black brane solution in $AdS_4$.  For this we have $\mathcal{E}=\frac{2 r_0^3}{16 \pi G_4 L}$, $T=\frac{3r_0}{4\pi L}$ and $s=\frac{4\pi r_0^2}{16 \pi G_4}$.  These satisfy the relations $d\mathcal{E}=T ds, \mathcal{E}=\frac{2}{3}Ts$.  The second relation agrees with our expression above for $z=1$.

In general, one may use 2 relations of the form
\bea
&&d\mathcal{E}=T ds \label{thermodef} \\
&& \mathcal{E}=K Ts \label{thermoRelgen}
\eea
($K$ a constant, which for our purposes is a function of $z$) to find the functional forms $\mathcal{E}(r_0)$, $\mathcal{E}(T)$ and $s(r_0)$, $s(T)$.  First, one may use (\ref{thermoRelgen}) to eliminate $ds$ in the relation (\ref{thermodef}), directly relating $dE$ and $dT$.  One may integrate this to find $\mathcal{E}(T)$.  One may then use this in (\ref{thermoRelgen}) to find $s(T)=\frac{4\pi r_0^2}{16 \pi G_4}$.  Such a procedure will furnish $T(r_0)$ and so we can find $\mathcal{E}(r_0)$ and $s(r_0)$.  Doing so, we find
\bea
s(r_0)&=&\frac{4\pi r_0^2}{16 \pi G_4}, \quad \mathcal{E}(r_0)=\Theta \left(\frac{K 4 \pi r_0^2}{16\pi G_4 \Theta }\right)^{\frac{1}{K}}, \quad T(r_0)=\left(\frac{K 4\pi r_0^2}{16\pi G_4 \Theta}\right)^{\frac{1-K}{K}}
\eea
where $\Theta$ is a constant of integration independent of $r_0$.  $\Theta$ has units of length$^{\frac{3K-2}{1-K}}$ so that the units of $s$ and $\mathcal{E}$ are canonical.  We can rewrite $\Theta$ in the following way.  $r_0$ is a metric parameter, and the temperature should only depend on metric parameters, not $16 \pi G_4$.  Further, the energy density should depend on $16 \pi G_4$, with one power of this factor in the denominator.  Therefore, we can say that $\Theta\propto \frac{1}{16 \pi G_4}$.  To make up the rest of the units, there is only one dimensionful parameter that one can use: $L$.  Therefore, we take that $\Theta = \frac{L^{\frac{K}{1-K}}}{16\pi G_4} n(z)$.  This gives
\bea
s(r_0)&=&\frac{4\pi r_0^2}{16 \pi G_4}, \qquad \qquad \qquad \qquad \qquad \qquad  s(T)=\frac{L^{\frac{K(z)}{1-K(z)}}}{16\pi G_4 K(z)} n(z) T^{\frac{K(z)}{1-K(z)}}, \nn \\
\mathcal{E}(r_0)&=&\frac{1}{16\pi G_4 L} n(z) \left(\frac{K(z) 4 \pi r_0^2}{n(z)}\right)^{\frac{1}{K(z)}}, \qquad \mathcal{E}(T)=\frac{L^{\frac{K(z)}{1-K(z)}}}{16\pi G_4} n(z) T^{\frac{1}{1-K(z)}},  \label{fcnsT} \\
T(r_0)&=&\frac{1}{L}\left(\frac{K(z) 4\pi r_0^2}{n(z)}\right)^{\frac{1-K(z)}{K(z)}} \nn
\eea
where we now explicitly write that $K$ is a function of $z$ $\left(K(z)=\frac{2}{2+z}\right)$.  To the right of each of these expressions, we have given the relation in terms of only the thermodynamic variable $T$, rather than referring to the geometric variable $r_0$.

One may use the above relations to show that the pressure $P$ is related to the energy density by
$2P=z\mathcal{E}$, which matches the holographic result of \cite{Ross:2009ar}.  Actually, starting from the equation of state $2P=z \mathcal{E}$ one can derive the relation (\ref{thermoRel}).

The value of $n(z)$, which is a unitless number, may be determined numerically by plotting $\log({LT})$ vs $\log({r_0})$,
but we did not attempt to do this here.

The slope of the graph of $\log(T(r_0))$ vs $\log(r_0)$ is
\be
\log(T(r_0))=\frac{2(1-K(z))}{K(z)}\log(r_0)+{\rm constant}=z\log(r_0)+{\rm constant},
\ee
which can be quantitatively checked using our earlier data \cite{Bertoldi:2009vn}.

We may easily compare the thermodynamic relationship \eqref{thermoRel} with the expected behavior from a system with a scaling symmetry $(x_1,x_2)\rightarrow \lambda (x_1,x_2)$, $t\rightarrow \lambda^z t$.  This scaling symmetry (along with $SO(2)$ symmetry rotating the $x_i$) implies that the dispersion relation is
\be
\omega^2=\alpha^2 (k_1^2+k_2^2)^z\equiv \alpha^2 k^{2z}
\ee
where $\alpha$ is some parameter to restore canonical units.   We will work with a finite system (a box with sides of length $\ell$), and assume that the occupation number of a given mode (with energy $E_n$) in the box is $\mathcal{Q}'\left(e^{-\beta E_n}\right)e^{-\beta E_n}$.  This way, the energy of the system is written as
\be
E=-\frac{\pa}{\pa \beta} Q, \qquad Q=\sum_n \mathcal{Q}\left(e^{-\beta E_n}\right).
\ee
Inside the box, $k_i=\frac{2\pi}{\ell}n_i$, and we have $d=2$ spatial dimensions, so there are two $n_i$.  We approximate the sum by an integral, and realize that the density of integer $(n_1,n_2)$ lattice points is uniform in the $(n_1,n_2)$ plane to find
\be
Q=2\pi\int n^2 \frac{dn}{n} \mathcal{Q}\left(e^{-\beta \alpha \left(\frac{2\pi}{\ell}n\right)^z}\right), \qquad n_1^2+n_2^2\equiv n^2.
\ee
Redefining the integration variable, we find
\be
\frac{\ell^2}{2 \pi z}\alpha^{-\frac2z}\beta^{-\frac2z}\int_0^\infty d\lambda \lambda^{\frac2z-1}\mathcal{Q}\left(e^{-\lambda}\right)\equiv \frac{z}{2} \Xi V \beta^{-\frac2z},
\ee
where the constant $\Xi$ is independent of $\beta$, and we have assumed that the function $\mathcal{Q}$ is well behaved at infinity (this is used to construct $\Xi$).  This yields all thermodynamic quantities
\be
E=\Xi V T^{\frac{2+z}{z}},\quad S=\int_{dV=0} dT \frac{1}{T} \left(\frac{dE}{dT}\right)_V=\Xi V T^{\frac2z}\frac{2+z}{2}
\ee
and so indeed relation (\ref{thermoRel}) holds, as well as all subsequent formulae.  Using the above argument, we can generalize the result to arbitrary spatial dimension $d$ where the relation (\ref{thermoRel}) becomes
\be
\mathcal{E}=Ts \frac{d}{d+z}. \label{thermoreld}
\ee
which generalizes the function $K(z)\rightarrow K(z,d)=\frac{d}{d+z}$ (we also promote $n(z) \rightarrow n(z,d)$).  For $d=3$ and $z=1$ (giving $K= \frac34$) one can check that (\ref{thermoreld}) and (\ref{fcnsT}) are correct by comparing to the known results for black D3 branes \cite{Gubser:1996de2} for the functions $s(T)$ and $\mathcal{E}(T)$ up to the normalization $n(z,d)$.  The functions of $s(r_0), \mathcal{E}(r_0), T(r_0)$ need to be modified by promoting $r_0^2
\rightarrow r_0^d$, so that the ``area'' is measured appropriately.  With this modification, the expressions for $s(r_0), \mathcal{E}(r_0), T(r_0)$ also agree; see for example \cite{Gubser:1998nz}, where similar coordinates are used.

\section*{Acknowledgements}

This work has been supported by NSERC of Canada.


\begin{thebibliography}{99}


\bibitem{Maldacena:1997re}
  J.~M.~Maldacena,
  %``The large N limit of superconformal field theories and supergravity,''
  Adv.\ Theor.\ Math.\ Phys.\  {\bf 2}, 231 (1998)
  [Int.\ J.\ Theor.\ Phys.\  {\bf 38}, 1113 (1999)],
  [arXiv:hep-th/9711200] %%CITATION = IJTPB,38,1113;%%;
  E.~Witten,
  %``Anti-de Sitter space and holography,''
  Adv.\ Theor.\ Math.\ Phys.\  {\bf 2}, 253 (1998)
  [arXiv:hep-th/9802150]. %%CITATION = 00203,2,253;%%


\bibitem{Aharony:1999ti}
  O.~Aharony, S.~S.~Gubser, J.~M.~Maldacena, H.~Ooguri and Y.~Oz,
  %``Large N field theories, string theory and gravity,''
  Phys.\ Rept.\  {\bf 323}, 183 (2000)
  [arXiv:hep-th/9905111].
  %%CITATION = PRPLC,323,183;%%

\bibitem{Gubser:1996de2}
  S.~S.~Gubser, I.~R.~Klebanov and A.~W.~Peet,
  %``Entropy and Temperature of Black 3-Branes,''
  Phys.\ Rev.\  D {\bf 54}, 3915 (1996)
  [arXiv:hep-th/9602135].
  %%CITATION = PHRVA,D54,3915;%%

\bibitem{Gubser:1996de}
  G.~Policastro, D.~T.~Son and A.~O.~Starinets,
  %``The shear viscosity of strongly coupled N = 4 supersymmetric Yang-Mills
  %plasma,''
  Phys.\ Rev.\ Lett.\  {\bf 87}, 081601 (2001)
  [arXiv:hep-th/0104066].
  %%CITATION = PRLTA,87,081601;%%
  A.~Buchel and J.~T.~Liu,
  %``Universality of the shear viscosity in supergravity,''
  Phys.\ Rev.\ Lett.\  {\bf 93}, 090602 (2004)
  [arXiv:hep-th/0311175].
  %%CITATION = PRLTA,93,090602;%%
  P.~Kovtun, D.~T.~Son and A.~O.~Starinets,
  %``Viscosity in strongly interacting quantum field theories from black hole
  %physics,''
  Phys.\ Rev.\ Lett.\  {\bf 94}, 111601 (2005)
  [arXiv:hep-th/0405231].
  %%CITATION = PRLTA,94,111601;%%
  A.~Buchel, J.~T.~Liu and A.~O.~Starinets,
  %``Coupling constant dependence of the shear viscosity in N=4 supersymmetric
  %Yang-Mills theory,''
  Nucl.\ Phys.\  B {\bf 707}, 56 (2005)
  [arXiv:hep-th/0406264].
  %%CITATION = NUPHA,B707,56;%%
  R.~C.~Myers, M.~F.~Paulos and A.~Sinha,
  %``Quantum corrections to eta/s,''
  Phys.\ Rev.\  D {\bf 79}, 041901 (2009)
  [arXiv:0806.2156 [hep-th]].
  %%CITATION = PHRVA,D79,041901;%%




\bibitem{Witten:1998zw}
  E.~Witten,
  %``Anti-de Sitter space, thermal phase transition, and confinement in  gauge
  %theories,''
  Adv.\ Theor.\ Math.\ Phys.\  {\bf 2}, 505 (1998)
  [arXiv:hep-th/9803131],
  %%CITATION = 00203,2,505;%%
  S.~W.~Hawking and D.~N.~Page,
  %``Thermodynamics Of Black Holes In Anti-De Sitter Space,''
  Commun.\ Math.\ Phys.\  {\bf 87}, 577 (1983).
  %%CITATION = CMPHA,87,577;%%

\bibitem{Hartnoll:2009sz}
  S.~A.~Hartnoll,
  %``Lectures on holographic methods for condensed matter physics,''
  arXiv:0903.3246 [hep-th].
  %%CITATION = ARXIV:0903.3246;%%
  C.~P.~Herzog,
  %``Lectures on Holographic Superfluidity and Superconductivity,''
  arXiv:0904.1975 [hep-th].
  %%CITATION = ARXIV:0904.1975;%%

\bibitem{Son:2008ye}
  D.~T.~Son,
  %``Toward an AdS/cold atoms correspondence: a geometric realization of the
  %Schroedinger symmetry,''
  Phys.\ Rev.\  D {\bf 78}, 046003 (2008)
  [arXiv:0804.3972 [hep-th]].
  %%CITATION = PHRVA,D78,046003;%%

\bibitem{Balasubramanian:2008dm}
K.~Balasubramanian and J.~McGreevy,
%``Gravity duals for non-relativistic CFTs,''
Phys.\ Rev.\ Lett.\  {\bf 101}, 061601 (2008)
[arXiv:0804.4053 [hep-th]].
%%CITATION = PRLTA,101,061601;%%

\bibitem{Mazzucato:2008tr}
  L.~Mazzucato, Y.~Oz and S.~Theisen,
  %``Non-relativistic Branes,''
  JHEP {\bf 0904}, 073 (2009)
  [arXiv:0810.3673 [hep-th]].
  %%CITATION = JHEPA,0904,073;%%
  J.~Maldacena, D.~Martelli and Y.~Tachikawa,
  %``Comments on string theory backgrounds with non-relativistic conformal
  %symmetry,''
  JHEP {\bf 0810}, 072 (2008)
  [arXiv:0807.1100 [hep-th]].
  %%CITATION = JHEPA,0810,072;%%
  %\cite{Bobev:2009mw}
  N.~Bobev, A.~Kundu and K.~Pilch,
  %``Supersymmetric IIB Solutions with Schr\'{o}dinger Symmetry,''
  arXiv:0905.0673 [hep-th].
  %%CITATION = ARXIV:0905.0673;%%
  W.~Y.~Wen,
  %``AdS/NRCFT for the (super) Calogero model,''
  arXiv:0807.0633 [hep-th].
  %%CITATION = ARXIV:0807.0633;%%


\bibitem{Yamada:2008if}
 A.~Adams, K.~Balasubramanian and J.~McGreevy,
  %``Hot Spacetimes for Cold Atoms,''
  JHEP {\bf 0811}, 059 (2008)
  [arXiv:0807.1111 [hep-th]].
  %%CITATION = JHEPA,0811,059;%%
  C.~P.~Herzog, M.~Rangamani and S.~F.~Ross,
  %``Heating up Galilean holography,''
  JHEP {\bf 0811}, 080 (2008)
  [arXiv:0807.1099 [hep-th]].
  %%CITATION = JHEPA,0811,080;%%
  D.~Yamada,
  %``Thermodynamics of Black Holes in Schroedinger Space,''
  Class.\ Quant.\ Grav.\  {\bf 26}, 075006 (2009)
  [arXiv:0809.4928 [hep-th]].
  %%CITATION = CQGRD,26,075006;%%
  E.~Imeroni and A.~Sinha,
  %``Non-relativistic metrics with extremal limits,''
  arXiv:0907.1892 [hep-th].
  %%CITATION = ARXIV:0907.1892;%%

\bibitem{Azeyanagi:2009pr}
  T.~Azeyanagi, W.~Li and T.~Takayanagi,
  %``On String Theory Duals of Lifshitz-like Fixed Points,''
  JHEP {\bf 0906}, 084 (2009)
  [arXiv:0905.0688 [hep-th]].
  %%CITATION = JHEPA,0906,084;%%


\bibitem{Kachru:2008yh}
  S.~Kachru, X.~Liu and M.~Mulligan,
  %``Gravity Duals of Lifshitz-like Fixed Points,''
  Phys.\ Rev.\  D {\bf 78}, 106005 (2008)
  [arXiv:0808.1725 [hep-th]].
  %%CITATION = PHRVA,D78,106005;%%

\bibitem{Taylor:2008tg}
  M.~Taylor,
  %``Non-relativistic holography,''
  arXiv:0812.0530 [hep-th].
  %%CITATION = ARXIV:0812.0530;%%

%\cite{Pang:2009ad}
\bibitem{Pang:2009ad}
 D.~W.~Pang,
 %``A Note on Black Holes in Asymptotically Lifshitz Spacetime,''
 arXiv:0905.2678 [hep-th].
 %%CITATION = ARXIV:0905.2678;%%

\bibitem{Kovtun:2008qy}
  P.~Kovtun and D.~Nickel,
  %``Black holes and non-relativistic quantum systems,''
  Phys.\ Rev.\ Lett.\  {\bf 102}, 011602 (2009)
  [arXiv:0809.2020 [hep-th]].
  %%CITATION = PRLTA,102,011602;%%
  
\bibitem{Koroteev:2007yp}
  P.~Koroteev and M.~Libanov,
  %``On Existence of Self-Tuning Solutions in Static Braneworlds without
  %Singularities,''
  JHEP {\bf 0802}, 104 (2008)
  [arXiv:0712.1136 [hep-th]].
  %%CITATION = JHEPA,0802,104;%%,
  P.~Koroteev and M.~Libanov,
  %``Spectra of Field Fluctuations in Braneworld Models with Broken Bulk Lorentz
  %Invariance,''
  Phys.\ Rev.\  D {\bf 79}, 045023 (2009)
  [arXiv:0901.4347 [hep-th]].
  %%CITATION = PHRVA,D79,045023;%%

\bibitem{Bertoldi:2009vn}
  G.~Bertoldi, B.~A.~Burrington and A.~Peet,
  %``Black Holes in asymptotically Lifshitz spacetimes with arbitrary critical
  %exponent,''
  arXiv:0905.3183 [hep-th].
  %%CITATION = ARXIV:0905.3183;%%

\bibitem{Danielsson:2008gi}
  U.~H.~Danielsson and L.~Thorlacius,
  %``Black holes in asymptotically Lifshitz spacetime,''
  arXiv:0812.5088 [hep-th].
  %%CITATION = ARXIV:0812.5088;%%

\bibitem{Mann:2009yx}
  R.~B.~Mann,
  %``Lifshitz Topological Black Holes,''
  arXiv:0905.1136 [hep-th].
  %%CITATION = ARXIV:0905.1136;%%

\bibitem{Ross:2009ar}
  S.~F.~Ross and O.~Saremi,
  %``Holographic stress tensor for non-relativistic theories,''
  arXiv:0907.1846 [hep-th].
  %%CITATION = ARXIV:0907.1846;%%

\bibitem{Gubser:1998nz}
  S.~S.~Gubser, I.~R.~Klebanov and A.~A.~Tseytlin,
  %``Coupling constant dependence in the thermodynamics of N = 4  supersymmetric
  %Yang-Mills theory,''
  Nucl.\ Phys.\  B {\bf 534}, 202 (1998)
  [arXiv:hep-th/9805156].
  %%CITATION = NUPHA,B534,202;%%



\end{thebibliography}
\end{document}